\documentclass{PoS}
\usepackage{cite}
\usepackage[latin1,utf8]{inputenc}

\newcommand{\bq}{\begin{eqnarray}}
\newcommand{\eq}{\end{eqnarray}}
\newcommand{\eps}{\varepsilon}

\title{Applications of intersection numbers in physics}

\ShortTitle{Applications of intersection numbers in physics}

\author{Stefan Weinzierl\\
        Johannes Gutenberg-Universit\"at Mainz\\
        E-mail: \email{weinzierl@uni-mainz.de}}

\abstract{
In this review I discuss intersection numbers of twisted cocycles and their relation to physics.
After defining what these intersection number are, I will first discuss a method for computing them.
This is followed by three examples where intersection numbers appear in physics.
These examples are: tree-level scattering amplitudes within the the CHY-formalism, reduction of Feynman integrals to master integrals
and correlation functions on the lattice.
}

\FullConference{MathemAmplitudes 2019: Intersection Theory \& Feynman Integrals\\18-20 December 2019\\	
Padova, Italy}

\begin{document}

\section{Introduction}

Intersection numbers of twisted cocycles are first of all a well-studied topic in mathematics
\cite{aomoto1975,Matsumoto:1994,cho1995,matsumoto1998,Ohara:2003,Goto:2013,Goto:2015aaa,Goto:2015aab,Goto:2015aac,Matsubara-Heo:2019,Aomoto:book,Yoshida:book,Matsubara-HeoPadova}.
Quite recently, it has been become clear that they are also relevant to physics \cite{Mizera:2017cqs,Mizera:2017rqa,Mastrolia:2018uzb,Weinzierl:2020nhw,Mizera:2020wdt}
and they provide an underlying mathematical framework for some established formulae and methods.
In this review we discuss how to compute these intersection numbers \cite{Mizera:2019gea,Frellesvig:2019uqt,Mizera:2019vvs,Weinzierl:2020xyy}
and give three examples, where intersection numbers occur in physics.
These examples are: 
(i) The Cachazo-He-Yuan formula \cite{Cachazo:2013gna,Cachazo:2013hca,Cachazo:2013iea,Mizera:2017cqs,Mizera:2017rqa,Mizera:2019gea,delaCruz:2017zqr}
for tree-level scattering amplitude,
(ii) the decomposition of Feynman integrals in terms of master integrals 
\cite{Mastrolia:2018uzb,Frellesvig:2019kgj,Frellesvig:2019uqt,Frellesvig:2020qot}
and
(iii) correlation functions on a lattice \cite{Weinzierl:2020nhw}.

\section{Intersection numbers}

Let us start by defining our objects of interest, i.e.
intersection numbers of twisted cocycles.
We first define cocycles, then twisted cocycles and finally intersection numbers for the latter.

We start from the $n$-dimensional complex space $\mathbb{C}^n$ and a divisor $D$ (i.e. a linear combination of sub-varieties of codimension one), 
on which we will allow singularities.
Let us assume that $D$ is defined by $m$ polynomial equations $p_i(z_1,\dots,z_n)=0$, $p_i \in \mathbb{C}\left[z_1,\dots,z_n\right]$
\bq
 D \; = \; 
 \bigcup\limits_{i=1}^m D_i
 \;\;\;
 & \mbox{and} & 
 \;\;\;
 D_i \; = \; \{ p_i = 0 \} \; \subset \; {\mathbb C}^n.
\eq
We consider rational differential $n$-forms $\varphi$ in the variables $z=(z_1,\dots,z_n)$, 
which are holomorphic on ${\mathbb C}^n - D$.
The rational $n$-forms $\varphi$ are of the form
\bq
\label{representative_left}
 \varphi
 & = &
 \frac{q}{p_1^{n_1} \dots p_m^{n_m}} \; dz_n \wedge \dots \wedge dz_1,
 \;\;\;\;\;\;\;\;\;
 q \in \mathbb{C}\left[z_1,\dots,z_n\right],
 \;\;\;
 n_i \in {\mathbb N}_0.
\eq
Using the reversed wedge product $dz_n \wedge \dots \wedge dz_1$ instead of the standard order $dz_1 \wedge \dots \wedge dz_n$
is at this stage just a convention.

In cohomology theory we call the differential $n$-form $\varphi$ a cocycle.
It is closed on ${\mathbb C}^n - D$, since it is a holomorphic $n$-form.
Let ${\mathcal C}$ be a $n$-dimensional integration cycle (i.e. an integration domain with no boundary $\partial {\mathcal C}=0$).
We may now consider the integral
\bq
 \left\langle \varphi | {\mathcal C} \right\rangle
 & = &
 \int\limits_{\mathcal C} \varphi.
\eq
This is a pairing between a cycle and a cocycle.
The quantity $\langle \varphi | {\mathcal C} \rangle$ will not change if we add to $\varphi$ the exterior derivative of
a holomorphic $(n-1)$-form $\xi$:
\bq
\label{invariance_d}
 \varphi & \rightarrow & \varphi + d\xi.
\eq
Due to $\partial {\mathcal C}=0$ and Stokes' theorem
\bq
 \int\limits_{\mathcal C} d\xi
 & = &
 \int\limits_{\partial \mathcal C} \xi
 \; = \; 0,
\eq
we have
\bq
 \left\langle \varphi + d\xi | {\mathcal C} \right\rangle
 & = &
 \left\langle \varphi | {\mathcal C} \right\rangle.
\eq
Let us now introduce the twist:
For $m$ complex numbers $\gamma=(\gamma_1,\dots,\gamma_m)$ 
we set
\bq
\label{def_u}
 u
 & = &
 \prod\limits_{i=1}^m p_i^{\gamma_i}.
\eq
Since the exponents $\gamma_i$ of the polynomials $p_i$ are allowed to be complex numbers,
$u$ is in general a multi-valued function on ${\mathbb C}^n - D$.
It will be convenient to define
\bq
 \omega
 & = &
 d \ln u
 \; = \;
 \sum\limits_{i=1}^m \gamma_i d\ln p_i
 \; = \;
 \sum\limits_{j=1}^n  \omega_j dz_j.
\eq
Let us fix a branch cut of $u$.
We then consider the integral
\bq
\label{def_integral}
 \left\langle \varphi | {\mathcal C} \right\rangle_{\omega}
 & = &
 \int\limits_{\mathcal C} u \varphi.
\eq
${\mathcal C}$ is again an integration cycle. 
We may allow ${\mathcal C}$ to have a boundary contained in $D$: $\partial {\mathcal C} \subset D$.
The integral remains well defined, if we assume that $\mathrm{Re}\gamma_i$ is sufficiently large, such that $u \varphi$ vanishes on $D$.
It is not too difficult to see that now the integral remains invariant under
\bq
\label{invariance_nabla_omega}
 \varphi & \rightarrow & \varphi + \nabla_\omega \xi,
\eq
where we introduced the covariant derivative $\nabla_\omega = d + \omega$.
In fact we have
\bq
 \int\limits_{\mathcal C} u \nabla_\omega \xi
 & = &
 \int\limits_{\mathcal C} \left[ u d \xi + u \left( d \ln u \right) \xi \right]
 \; = \;
 \int\limits_{\mathcal C} d \left( u \xi \right)
 \; = \;
 \int\limits_{\partial \mathcal C} u \xi
 \; = \; 0.
\eq
Introducing the twist amounts to going from the normal derivative $d$ in eq.~(\ref{invariance_d}) 
to the covariant derivative $\nabla_\omega = d + \omega$ in eq.~(\ref{invariance_nabla_omega}).
The invariance under eq.~(\ref{invariance_nabla_omega})
motivates the definition of equivalence classes of $n$-forms $\varphi$: 
Two $n$-forms $\varphi'$ and $\varphi$ are called equivalent, if they differ by a covariant derivative
\bq
 \varphi'
 \sim
 \varphi
 & \;\; \Leftrightarrow \;\; &
 \varphi'
 \; = \;
  \varphi
  + \nabla_\omega \xi
\eq
for some $(n-1)$-form $\xi$.
We denote the equivalence classes by $\langle \varphi |$.
Being $n$-forms, each $\varphi$ is closed with respect to $\nabla_\omega$ and the equivalence classes
define the twisted cohomology group $H^n_\omega$:
\bq
 \left\langle \varphi \right|
 & \in &
 H^n_\omega.
\eq
The dual twisted cohomology group is given by
\bq
 \left( H^n_\omega \right)^\ast 
 & = &
 H^n_{-\omega}.
\eq
Elements of $( H^n_\omega )^\ast$ are denoted by $| \varphi \rangle$.
We have
\bq
 \left| \varphi' \right\rangle
 =
 \left| \varphi \right\rangle
 & \;\; \Leftrightarrow \;\; &
 \varphi'
 \; = \;
  \varphi
  + \nabla_{-\omega} \xi
\eq
for some $(n-1)$-form $\xi$.
A representative of a dual cohomology class is of the form
\bq
\label{representative_right}
 \varphi
 & = &
 \frac{q}{p_1^{n_1} \dots p_m^{n_m}} \; dz_1 \wedge \dots \wedge dz_n,
 \;\;\;\;\;\;\;\;\;
 q \in {\mathbb C}\left[z_1,\dots,z_n\right],
 \;\;\;
 n_i \in {\mathbb N}_0.
\eq
It will be convenient to use here the order $dz_1 \wedge \dots \wedge dz_n$ in the wedge product.

There is a non-degenerate bilinear pairing between a cohomology class $\langle \varphi_L|$ 
and a dual cohomology class $| \varphi_R \rangle$, given by the intersection number
\bq
 \left\langle \varphi_L \right. \left| \varphi_R \right\rangle_\omega.
\eq
The intersection number is defined by \cite{cho1995,Aomoto:book}
\bq
\label{def_intersection_number}
 \left\langle \varphi_L \right. \left| \varphi_R \right\rangle_\omega
 & = &
 \frac{1}{\left(2\pi i\right)^n}
 \int \iota_\omega\left(\varphi_L\right) \wedge \varphi_R
 \; = \;
 \frac{1}{\left(2\pi i\right)^n}
 \int \varphi_L \wedge \iota_{-\omega}\left(\varphi_R\right),
\eq
where $\iota_\omega$ maps $\varphi_L$ to its compactly supported version,
and similar for $\iota_{-\omega}$.
Please note that the pairing $\langle \varphi | {\mathcal C} \rangle_{\omega}$ 
between an integrand and an integration contour 
denotes the integral defined in eq.~(\ref{def_integral}),
while the pairing $\langle \varphi_L | \varphi_R \rangle_\omega$ between an integrand and a dual integrand 
denotes the intersection number defined in eq.~(\ref{def_intersection_number}).
Computing the intersection number through the definition in eq.~(\ref{def_intersection_number})
is not the most practical way (but see the appendix of ref.~\cite{Mizera:2017rqa} and ref.~\cite{CaronHuotPadova} in these proceedings) 
and we seek alternative methods to compute the intersection numbers.

\section{Computation of intersection numbers}

Let us now turn to the computation of a multivariate intersection numbers.
The word ``multivariate'' refers to the fact that for $n>1$ we have several variables $z_1, \dots, z_n$.
With a few technical assumptions, outlined in \cite{Mizera:2019gea,Frellesvig:2019uqt,Weinzierl:2020xyy}
we may compute multivariate intersection numbers in $n$ variables $z_1, \dots z_n$ recursively by splitting
the problem into the computation of an intersection number in $(n-1)$ variables $z_1, \dots, z_{n-1}$
and the computation of a (generalised) intersection number in the variable $z_n$.
By recursion, we therefore have to compute only (generalised) intersection numbers in a single variable $z_i$.
This reduces the multivariate problem to an univariate problem.

Let us comment on the word ``generalised'' intersection number:
We only need to discuss the univariate case.
Consider two cohomology classes $\langle \varphi_L |$ and $| \varphi_R \rangle$.
Representatives $\varphi_L$ and $\varphi_R$ for the two cohomology classes
$\langle \varphi_L |$ and $| \varphi_R \rangle$ are in the univariate case differential one-forms and of the form as
in eq.~(\ref{representative_left}) or eq.~(\ref{representative_right}).
We may view the representatives $\varphi_L$ and $\varphi_R$,
the cohomology classes $\langle \varphi_L |$ and $| \varphi_R \rangle$, and the twist $\omega$ 
as scalar quantities.

Consider now a vector of $\nu$ differential one-forms $\varphi_{L,j}$ in the variable $z$, 
where $j$ runs from $1$ to $\nu$.
Similar, consider for the dual space a $\nu$-dimensional vector $\varphi_{R,j}$
and generalise $\omega$ to a $(\nu\times \nu)$-dimensional matrix $\Omega$.
The equivalence classes $\langle \varphi_{L,j} |$ and $| \varphi_{R,j} \rangle$ are now defined
by
\bq
\label{invariance_Omega}
 \varphi_{L,j}' \; = \; \varphi_{L,j} + \partial_z \xi_j + \xi_i \Omega_{i j}
 & \;\; \mbox{and} \;\; &
 \varphi_{R,j}' \; = \; \varphi_{R,j} + \partial_z \xi_j - \Omega_{j i} \xi_i,
\eq
for some zero-forms $\xi_j$ (i.e. functions).
Readers familiar with gauge theories will certainly recognise that the generalisation is exactly the same
step as going from an Abelian gauge theory (like QED) to a non-Abelian gauge theory (like QCD).

The generalised intersection numbers sneak in as follows:
Let us set
\bq
 \omega^{({\bf i})}
 \; = \;
 \sum\limits_{j=1}^i \omega_j dz_j,
 & &
 \;\;\;
 H^{({\bf i})}_\omega
 \; = \;
 H^{i}_{\omega^{({\bf i})}}.
\eq
For example, $H^{({\bf n-1})}_\omega$ denotes the twisted cohomology 
group of $(n-1)$-forms in the variables $z_1, \dots, z_{n-1}$, where the remaining variable $z_n$ is treated as a parameter.
In other words, classes in $H^{({\bf n-1})}_\omega$ are represented by a rational function in $z_1, \dots, z_n$ times 
$dz_{n-1} \wedge \dots \wedge dz_1$.
The essential step in the recursive approach is to expand 
the twisted cohomology class $\langle \varphi_L | \in H^{({\bf n})}_\omega$ 
in the basis of $H^{({\bf n-1})}_\omega$:
\bq
 \left\langle \varphi_L \right|
 & = &
 \sum\limits_{j=1}^{\nu}
 \left\langle \varphi_{L,j} \right| \wedge \left\langle e_j \right|.
\eq
Here, $\langle e_j |$ denotes a basis of $H^{({\bf n-1})}_\omega$.
The coefficients $\langle \varphi_{L,j}|$ are one-forms proportional to $dz_n$.
The invariance of the original class $\langle \varphi_L |$ under a transformation as in eq.~(\ref{invariance_nabla_omega})
translates into the invariance of the vector of coefficients $\langle \varphi_{L,j}|$
as in eq.~(\ref{invariance_Omega}).

The algorithm for computing a multivariate intersection number
consists of three steps:
\begin{enumerate}
\item Recursive approach: The algorithm integrates out one variable at a time.
This part has been outlined above.
It has the advantage to reduce a multivariate problem to a univariate problem.
\item Reduction to simple poles: In general we deal in cohomology with equivalence classes.
We may replace a representative
of an equivalence class with higher poles with an equivalent representative
with only simple poles.
This is similar to integration-by-part reduction.
However, let us stress that the involved systems of linear
equations are usually significantly smaller compared to standard integration-by-part reduction.
\item Evaluation of the intersection number as a global residue. Having reduced our objects to simple poles,
we may evaluate the intersection in one variable as an univariate global residue.
This is easily computed
and does not involve algebraic extensions like square roots.
\end{enumerate}
This algorithm exploits the fact that for representatives $\varphi_{L,j}$ and $\varphi_{R,j}$ which only have simple poles
the intersection number
\bq
 \left\langle \varphi_L \right. \left| \varphi_R \right\rangle_\omega
 & = &
 \sum\limits_{j=1}^{\nu_{\bf n-1}}
 \left\langle \varphi_{L,j} \left| \varphi_{R,j} \right. \right\rangle_\Omega
\eq
is a global residue and can be computed without introducing algebraic extensions.
The method to do this is an adaption of ref.~\cite{Cattani:2005,Sogaard:2015dba} to the univariate case.

\section{Example 1: The CHY-formalism}

Let us now turn to some examples from physics.
We start with the CHY-formalism \cite{Cachazo:2013gna,Cachazo:2013hca,Cachazo:2013iea}.
For a review of the CHY-formalism see \cite{Weinzierl:2016bus}.
The CHY-formalism allows us to express tree-level $n$-point amplitudes from three different theories 
in terms of two ``half-integrands''.
The amplitudes from the three theories are:
(i) double-cyclic-ordered amplitudes $m_{n}(\sigma, \tilde{\sigma}, p)$ in bi-adjoint scalar theory,
(ii) cyclic-ordered amplitudes $A_n(\sigma,p,\eps)$ in Yang-Mills theory
(iii) unordered amplitudes $M_n(p,\eps,\tilde{\eps})$ in perturbative gravity.
Here, we denote the $n$-tuple of external momenta by $p=(p_1,\dots,p_n)$,
cyclic orders by permutations $\sigma=(\sigma_1,...,\sigma_n)$ and $\tilde{\sigma}=(\tilde{\sigma}_1,\dots,\tilde{\sigma}_n)$,
and $n$-tuples of external polarisations by $\eps=(\eps_1,...,\eps_n)$ and $\tilde{\eps}=(\tilde{\eps}_1,\dots,\tilde{\eps}_n)$.
One further denotes by $z=(z_1,\dots,z_n)$ a $n$-tuple of auxiliary complex variables.
The amplitude $M_n(p,\eps,\tilde{\eps})$ describes the scattering of $n$ gravitons. Being spin $2$ particles, the polarisation of a graviton
is described by a product $\eps_j \tilde{\eps}_j$ of two polarisation vectors.
The CHY-representation for these amplitudes reads
\bq
\label{CHY_representation}
 m_{n}\left( \sigma, \tilde{\sigma}, p \right)
 & = &
 i \oint\limits_{\mathcal C} d\Omega_{\mathrm{CHY}} \; 
 C\left(\sigma,z\right) \; C\left(\tilde{\sigma},z\right),
 \nonumber \\
 A_n\left(\sigma,p,\eps\right) & = &
 i \oint\limits_{\mathcal C} d\Omega_{\mathrm{CHY}} \; 
 C\left(\sigma,z\right) \; E\left(p,\eps,z\right),
 \nonumber \\
 M_n\left(p,\eps,\tilde{\eps}\right) & = &
 i \oint\limits_{\mathcal C} d\Omega_{\mathrm{CHY}} \; 
 E\left(p,\eps,z\right) \; E\left(p,\tilde{\eps},z\right),
\eq
where the integration contour encircles the inequivalent solutions of the scattering equations
\bq
 f_i\left(z,p\right) & = & 
 \sum\limits_{j=1, j \neq i}^n \frac{ 2 p_i \cdot p_j}{z_i - z_j}
 \; = \;
 0,
 \;\;\;\;\;\;\;\;\;
 1\le i \le n.
\eq
The measure is given by
\bq
 d\Omega_{\mathrm{CHY}}
 & = &
 \frac{1}{\left(2\pi i\right)^{n-3}}
 \frac{d^nz}{d\eta}
 \;
 \prod{}' \frac{1}{f_a\left(z,p\right)},
 \nonumber \\
 & &
 \prod{}' \frac{1}{f_a\left(z,p\right)}
 \; = \; 
 \left(-1\right)^{i+j+k}
 \left( z_i - z_j \right) \left( z_j - z_k \right) \left( z_k - z_i \right)
 \prod\limits_{a \neq i,j,k} \frac{1}{f_a\left(z,p\right)},
 \nonumber \\
 & &
 d\eta
 \; = \;
 \left(-1\right)^{p+q+r}
 \frac{dz_p dz_q dz_r}{\left( z_p - z_q \right) \left( z_q - z_r \right) \left( z_r - z_q \right)}.
\eq
The most important ingredients are the two ``half-integrands'', the cyclic factor $C(\sigma,z)$
and the polarisation factor $E(p,\eps,z)$.
The cyclic factor (or Parke-Taylor factor) 
is given by
\bq
 C\left(\sigma,z\right)
 & = & 
 \frac{1}{\left(z_{\sigma_1} - z_{\sigma_2} \right) \left( z_{\sigma_2} - z_{\sigma_3} \right) ... \left( z_{\sigma_n} - z_{\sigma_1} \right)}
\eq
and encodes the information on the cyclic order. 
The polarisation factor $E\left(p,\eps,z\right)$ 
encodes the information on the helicities of the external particles.
One possibility to define this factor is through a reduced Pfaffian \cite{Cachazo:2013hca}.
However, the integral representation in eq.~(\ref{CHY_representation}) localises the integrand on the solutions of the scattering equations
and hence only the value on the solutions of the scattering equations matters.
We may therefore redefine the polarisation factor such that this factor agrees with the reduced Pfaffian
on the solutions of the scattering equations, but is allowed to differ away from this zero-dimensional sub-variety.

S{\o}gaard and Zhang have shown that the CHY-representation is a global residue \cite{Sogaard:2015dba,Bosma:2016ttj}.
This allows us to evaluate the integrals in eq.~(\ref{CHY_representation}) 
without the need to know the solutions of the scattering equations.
This is advantageous, as summing the individual residues from the solutions of the scattering equations will in general introduce
algebraic extensions (e.g. square roots) \cite{Weinzierl:2014vwa}.
In the end, the algebraic extensions will drop out, if the sum over all residues is taken.

The scattering equations are invariant under the projective special linear group
$\mathrm{PSL}(2,{\mathbb C})$.
This allows us to fix three of the $n$ auxiliary variables $z_j$ at prescribed values, typically $0$, $1$ and $\infty$.
We will refer to this procedure (fixing three variables at prescribed values) as gauge-fixing.
In mathematical terms the auxiliary space of the $z$-variables is just
the moduli space ${\mathcal M}_{0,n}$ of an algebraic curve of genus zero (i.e. a Riemann sphere) with $n$ distinct marked points:
\bq
 {\mathcal M}_{0,n}
 & = &
 \left\{ z \in \left( {\mathbb C} {\mathbb P}^1 \right)^n : z_i \neq z_j \right\}/\mathrm{PSL}\left(2,{\mathbb C}\right).
\eq
${\mathcal M}_{0,n}$ is an affine algebraic variety of dimension $(n-3)$.

Let us now make contact with intersection theory. 
To this aim we define $(n-3)$-forms
by
\bq
 \Omega^{\mathrm{cyclic}}\left(\sigma,z\right)
 \;= \;
  C\left(\sigma,z\right) \frac{d^nz}{d\eta},
 & \;\;\; &
 \Omega^{\mathrm{pol}}\left(p,\eps,z\right)
 \; = \;
  E\left(p,\eps,z\right) \frac{d^nz}{d\eta}
\eq
and a connection one-form by
\bq
 \omega & = &
 \left. \sum\limits_{i=1}^n f_i\left(z,p\right) dz_i \right|_{\mathrm{gauge-fixed}}.
\eq
$\omega$ correspond to the gauge-fixed differential of the Koba-Nielsen function
\bq 
 u\left(z,p\right)
 & = &
 \prod\limits_{i<j} \left(z_i-z_j\right)^{2p_i \cdot p_j}.
\eq
We call $\Omega^{\mathrm{cyclic}}$ and $\Omega^{\mathrm{pol}}$ scattering forms.
If the scattering forms have only simple poles, the intersection number is given by the global residue.
We already mentioned above that the global residue equals the scattering amplitude.
We therefore conclude that in the case where the scattering forms have only simple poles, their intersection number
equals the scattering amplitude.

It is clear from the explicit expression for $C(\sigma,z)$ that $\Omega^{\mathrm{cyclic}}$ has only simple poles.
However this is not true for $\Omega^{\mathrm{pol}}$, if we define $E(p,\eps,z)$ through the reduced Pfaffian.
The reduced Pfaffian will in general lead to 
higher poles \cite{Du:2013sha,Litsey:2013jfa,Lam:2016tlk,Bjerrum-Bohr:2016axv,Huang:2017ydz,Du:2017kpo,Gao:2017dek,Chen:2016fgi,Chen:2017edo,Chen:2017bug}.
However, it is possible to define $E(p,\eps,z)$ such that it agrees with the reduced Pfaffian on the solutions of the scattering equations
and has only simple poles.
This is the ``comb'' representation of the polarisation factor \cite{delaCruz:2017zqr,Tolotti:2013caa}
\bq
\label{def_polarisation_factor}
 E\left(p,\eps,z\right)
 & = &
 \sum\limits_{\sigma \in S_{n-2}^{(1,n)}}
 C\left(\sigma,z\right)
 \;
 N_{\mathrm{comb}}\left(\sigma\right),
\eq
where the sum is now over all permutations keeping $\sigma_1=1$ and
$\sigma_n=n$ fixed. The definition of $N_{\mathrm{comb}}\left(\sigma\right)$ is given in ref.~\cite{delaCruz:2017zqr}.
With the definition of the polarisation factor as in eq.~(\ref{def_polarisation_factor})
the scattering amplitudes are then given as intersection numbers \cite{Mizera:2017cqs,Mizera:2017rqa,Mizera:2019gea,delaCruz:2017zqr}
\bq
 m_{n}\left( \sigma, \tilde{\sigma}, p \right)
 & = &
 i \left\langle \Omega^{\mathrm{cyclic}}\left(\sigma,z\right), \Omega^{\mathrm{cyclic}}\left(\tilde{\sigma},z\right) \right\rangle_\omega,
 \nonumber \\
 A_n(\sigma,p,\eps)
 & = &
 i \left\langle \Omega^{\mathrm{cyclic}}\left(\sigma,z\right), \Omega^{\mathrm{pol}}\left(p,\eps,z\right) \right\rangle_\omega,
 \nonumber \\
 M_n(p,\eps,\tilde{\eps})
 & = &
 i \left\langle \Omega^{\mathrm{pol}}\left(p,\eps,z\right), \Omega^{\mathrm{pol}}\left(p,\tilde{\eps},z\right) \right\rangle_\omega.
\eq
Let us stress that 
the Cachazo-He-Yuan formula in eq.~(\ref{CHY_representation}) 
is always a global residue. If both half-integrand have simple poles, it is also an intersection number.

\section{Example 2: Feynman integrals}

As a second example we consider Feynman integrals and here in particular the reduction to master integrals.
Our starting point is a $l$-loop $N$-point Feynman integral
\bq
 I_{\nu_1 \nu_2 \dots \nu_{n}} & = &
 \int \prod\limits_{r=1}^l \frac{d^Dk_r}{i \pi^{\frac{D}{2}}}
 \prod\limits_{s=1}^{n}
 \frac{1}{\left( -q_s^2 + m_s^2 \right)^{\nu_s}},
 \;\;\;\;\;\;
 \nu_s \; \in \; {\mathbb Z}.
\eq
We are interested in the family of Feynman integrals indexed by $\nu_1,\nu_2,\dots,\nu_n$.
It is well-known that all integrals from this family may always been written as a linear combination 
of master integrals.
The standard tool to achieve this are integration-by-parts identities \cite{Tkachov:1981wb,Chetyrkin:1981qh}
\bq
 \int 
 \prod\limits_{r=1}^l \frac{d^Dk_r}{i \pi^{\frac{D}{2}}}
 \;\;
 \frac{\partial}{\partial k_i^\mu} \; a^\mu 
 \;\;
 \prod\limits_{s=1}^{n}
 \frac{1}{\left( -q_s^2 + m_s^2 \right)^{\nu_s}}
 & = & 0,
 \;\;\;\;\;\;\;\;\;
 a \in \{p_1,...,p_N,k_1,...,k_l\}
\eq
Working out the derivatives leads to relations among integrals 
with different sets of indices $\nu_1,\dots,\nu_n$.
These relations are then solved systematically \cite{Laporta:2001dd}, allowing us to express any integral
in terms of a linear combination of master integrals.

Alternatively, we may perform the reduction to master integrals 
through intersection numbers \cite{Mastrolia:2018uzb,Frellesvig:2019kgj,Frellesvig:2019uqt,Frellesvig:2020qot,Abreu:2019wzk,Chen:2020uyk}.
This by-passes the need to solve large systems of linear equations.
This is most easily seen in the Baikov representation \cite{Baikov:1996iu,Lee:2009dh}.
Let $p_1, p_2, ..., p_N$ denote the external momenta
and denote by
\bq
 e & = &
 \dim \left\langle p_1, p_2, ..., p_N \right\rangle
\eq
the dimension of the span of the external momenta.
For generic external momenta and $D \ge N-1$ we have $e=N-1$.
In order to arrive at the Baikov representation 
we change the integration variables to 
the Baikov variables $z_s$:
\bq
 z_s & = & -q_s^2+m_s^2.
\eq
The Baikov representation of the Feynman integral is given (schematically) by
\bq
\label{baikov_representation}
 I_{\nu_1 \dots \nu_{n}} & = &
 C
 \int\limits_{\mathcal C} d^{n}z \;
 B\left(z_1,...,z_{n}\right)^{\frac{D-l-e-1}{2}}
 \prod\limits_{s=1}^{n} z_s^{-\nu_s}.
\eq
$C$ is a prefactor and not relevant for the further discussion.
$B(z_1,...,z_{n})$ denotes the Baikov polynomial. It is obtained from a Gram determinant.
The domain of integration is such that the Baikov polynomial vanishes on the boundary of the integration region.
We note that the indices $\nu_s$ enter only the last factor.
Eq.~(\ref{baikov_representation}) is an integral of the form as in eq.~(\ref{def_integral})
with
\bq
 \varphi & = & \left( \prod\limits_{s=1}^{n} z_s^{-\nu_s} \right) d^{n}z 
\eq
and
\bq
 u & = & B\left(z_1,...,z_{n}\right)^{\frac{D-l-e-1}{2}},
 \;\;\;\;\;\;
 \omega \; = \; d \ln u.
\eq
As the Baikov polynomial vanishes on the boundary of the integration region, the Feynman integral is invariant
under
\bq
 \varphi & \rightarrow & \varphi + \nabla_\omega \xi
\eq
and we may group the integrands of the Feynman integrals $I_{\nu_1 \dots \nu_n}$ corresponding to different sets of indices $\nu_1, \dots, \nu_n$
into cohomology classes.
The number of independent cohomology classes in $H_\omega^n$ is finite, and we may express any $\varphi$ as a linear combination
of a basis of $H_\omega^n$.
Let $\langle e_j |$ be a basis of $H_\omega^n$ and $| d_j \rangle$ a basis of the dual cohomology group
$( H^n_\omega )^\ast$, chosen such that
\bq
 \left\langle e_i | d_j \right\rangle_\omega & = & \delta_{ij}.
\eq
We then have
\bq
 \left\langle \varphi \right|
 & = &
 \sum\limits_j c_j \left\langle e_j \right|,
\eq
where the coefficients are given by the intersection numbers
\bq
 c_j & = & 
 \left\langle \varphi | d_j \right\rangle_\omega.
\eq
This provides an alternative to integration-by-parts reduction.
Note that the dimension of $H_\omega^n$ can be larger than the number of master integrals, as the latter takes symmetries of integrals
into account, while the former operates on integrands.
This is most easily explained by the simplest example, the one-loop two-point function with two equal internal masses.
This system has two master integrals. A standard choice is $I_{11}$ and $I_{10}$.
By symmetry, the integral $I_{01}$ is identical to $I_{10}$.
At the level of the integrands we have $\dim H_\omega^2=3$.
A basis for $H_\omega^2$ is given by $\langle \varphi_{11} |$, $\langle \varphi_{10} |$, $\langle \varphi_{01} |$
with $\varphi_{\nu_1 \nu_2} = d^2z/(z_1^{\nu_1} z_2^{\nu_2})$. 
The $2$-forms $\varphi_{10}$ and $\varphi_{01}$ are not identical 
(of course, one is obtained from the other up to a sign through the substitution $z_1 \leftrightarrow z_2$),
only the integrals as in eq.~(\ref{def_integral}) give identical results.

\section{Example 3: Correlation functions on the lattice}

As our last example we consider correlation function on a lattice \cite{Weinzierl:2020nhw}.
Let us consider a lattice $\Lambda$ with lattice spacing $a$ and $L$ points in each direction.
In four space-time dimensions the lattice has then $N=L^4$ lattice points.
A lattice point is specified by a $4$-tuple
\bq
 x & = &
 \left( j_0, j_1, j_2, j_3 \right),
 \;\;\;\;\;\;
 0 \; \le \; j_i \; < \; L.
\eq
One of the simplest examples is scalar $\phi^4$-theory,
with the discretised Euclidean action
\bq
 S_E
 & = &
 \sum\limits_{x \in \Lambda}
 \left( 
 -  \sum\limits_{\mu=0}^{3} \phi_x \phi_{x+a e_\mu}
 + 4 \phi_x^2
 + \frac{\lambda}{4!} \phi_x^4
 \right).
\eq
$\phi_x$ denotes the field at the lattice point $x$ and
$\phi_{x+a e_\mu}$ denotes the field at the next lattice point in the (positive) $\mu$-direction modulo $L$.
Typically, one is interested in correlation functions like
\bq
\label{lattice_integral}
 I{\footnotesize \left(\begin{array}{ccc} \nu_1, & \dots, & \nu_n \\ x_1, & \dots, & x_n \end{array} \right)}
 & = &
\int\limits_{{\mathcal C}^N} d^N\phi \; \phi_{x_1}^{\nu_1} \dots \phi_{x_n}^{\nu_n} \; \exp\left(-S_E\right),
 \;\;\;\;\;\;
 \nu_j \; \in \; {\mathbb N}_0.
\eq
This is a finite-dimensional integral and of the form as eq.~(\ref{def_integral}).
This becomes evident if we define a function $u$, a one-form $\omega$ and a $N$-form $\varphi$ by
\bq
\label{def_twisted}
 u & = & \exp\left(-S_E\right),
 \nonumber \\
 \omega & = & d \ln u \;= \; -d S_E,
 \nonumber \\
 \varphi & = & \phi_{x_1}^{\nu_1} \dots \phi_{x_n}^{\nu_n} \; d^N\phi.
\eq
In terms of these quantities, the integral in eq.~(\ref{lattice_integral}) can be written as
\bq
 I{\footnotesize \left(\begin{array}{ccc} \nu_1, & \dots, & \nu_n \\ x_1, & \dots, & x_n \end{array} \right)}
 & = &
 \int\limits_{{\mathcal C}^N} u \; \varphi.
\eq
As each $\nu_j$ can take any value in ${\mathbb N}_0$, there are infinitely many correlation functions.
However, they are not independent.
The integrand is invariant under
\bq
\label{ibp_lattice}
 \varphi & \rightarrow & \varphi + \nabla_\omega \xi
\eq
and we may express any correlation function as a linear combination of correlation functions with integrands forming
a basis of $H^N_\omega$.
For the case of $\phi^4$-theory the dimension of $H^N_\omega$ is given by
\bq
 \dim H^N_\omega
 & = & 
 3^N.
\eq
A basis of $H^N_\omega$ is given by
\bq
\label{def_basis}
 \left\langle e_i \right| \; : \;\;\;
 \left( \prod\limits_{k=1}^N \phi_{x_k}^{\nu_k} \right) d^N\phi,
 & &
 \;\;\;
 0  \; \le \; \nu_k \; \le \;  2.
\eq
Let $| d_j \rangle$ denote a basis of the dual cohomology group
$( H^n_\omega )^\ast$ such that
\bq
 \left\langle e_i | d_j \right\rangle_\omega & = & \delta_{ij}.
\eq
We then have
\bq
 \left\langle \varphi \right|
 & = &
 \sum\limits_j c_j \left\langle e_j \right|,
\eq
where the coefficients are given by the intersection numbers
\bq
 c_j & = & 
 \left\langle \varphi | d_j \right\rangle_\omega.
\eq
Instead of first constructing the dual basis $| d_j \rangle$ and then computing intersection numbers, one could alternatively
reduce $\langle \varphi |$ to a basis by exploiting eq.~(\ref{ibp_lattice}) to reduce systematically the exponents $\nu_j$.
This alternative method is analogous to integration-by-parts reduction for Feynman integrals.

\section{Conclusion}

Intersection theory provides a mathematical framework to describe several -- at first sight unrelated situations -- in physics.
We discussed tree-level scattering amplitudes within the the CHY-formalism, 
reduction of Feynman integrals to master integrals
and correlation functions on the lattice.
Further development of intersection theory, tailored to the needs of physics and here in particular to methods
to compute efficiently intersection numbers will have an impact on application in physics.
The most promising application will be the reduction of Feynman integrals to master integrals.

\subsection*{Acknowledgements}

I would like to thank the organisers of the workshop
"MathemAmplitudes 2019: Intersection Theory and Feynman Integrals" 
for setting up a stimulating and productive meeting.

\bibliography{/home/stefanw/notes/biblio}
\bibliographystyle{/home/stefanw/latex-style/h-physrev5}

\end{document}